# Rigidity analysis of HIV-1 protease

J W Heal[1], S A Wells[2], E Jimenez-Roldan[2], R F Freedman[3] and R A Römer[2]

[1]MOAC Doctoral Training Centre, University of Warwick, Coventry, CV4 7AL, United Kingdom
[2]Department of Physics and Centre for Scientific Computing, University of Warwick, Coventry, CV4 7AL, United Kingdom
[3]School of Life Sciences, University of Warwick, Coventry, CV4 7AL, United Kingdom.

E-mail: jack.heal@warwick.ac.uk

**Abstract**

We present a rigidity analysis on a large number of X-ray crystal structures of the enzyme HIV-1 protease using the 'pebble game' algorithm of the software FIRST. We find that although the rigidity profile remains similar across a comprehensive set of high resolution structures, the profile changes significantly in the presence of an inhibitor. Our study shows that the action of the inhibitors is to restrict the flexibility of the β-hairpin flaps which allow access to the active site. The results are discussed in the context of full molecular dynamics simulations as well as data from NMR experiments.

## 1. Introduction

The relationships between protein structure, rigidity, and function are important in biophysics. In this study the software FIRST is used to create rigidity profiles of a set of 280 X-ray crystal structures of HIV-1 protease that have been solved to a resolution of at least 2.0 Å and made available via the protein data bank (PDB). FIRST runs the pebble game algorithm [1] to identify rigid clusters and flexible regions within a protein by matching atomic degrees of freedom with the constraints present due to bonding, and can carry out the rigidity analysis without high computational cost. By rapidly extracting information from all of the available high-resolution HIV-1 protease structures we use FIRST to build an overview of the information available in the PDB. In a recent study [2], FIRST is applied to a set of individual structures from each of six different proteins, one of which is HIV-1 protease. Here we examine the enzyme in more detail by making a comprehensive study of the high-resolution structures available.

Molecular dynamics (MD) studies on structures of HIV-1 protease [3] are informative yet computationally intensive and as such cannot offer an overview of the available structures. Experimentally, NMR [4] and fluorescence spectroscopy [5] have been used to study the flexibility of the protein. HIV-1 protease is a well-documented drug target due to its role in the life-cycle of HIV-1 [6] and many inhibitors have been designed in an attempt to block HIV-1 replication [7]. The structures have a range of crystallisation conditions, inhibitors and mutations. Initial findings from the analysis are presented, and work is ongoing to determine other underlying trends in the rigidity data of the 280 structures.

## 2. Method

In order to prepare the structures for rigidity analysis, all the water molecules are removed and the REDUCE software is used to add the hydrogen atoms that are not present in the X-ray crystal structure and to flip side chains where necessary. Uninhibited protease structures are created in PYMOL by manually by deletion of the inhibitor. A new PDB file is created representing the protein in a conformation which is part of the ensemble explored by the protein in its natural flexible motion [8]. The MD study in [9] also used this method to get structures of the apoenzyme and the holoenzyme. Covalent bonds, hydrogen bonds, salt bridges and hydrophobic interactions are included as constraints that restrict the freedom of the atoms in a molecule. Weaker, longer-range electrostatic forces and van der Waals forces are omitted [10]. FIRST builds a rigidity profile of the protein consisting of rigid substructures linked together by flexible sections [11]. The profile is calculated at a particular energy cutoff value Ecut, and whilst other constraints are maintained, only hydrogen bonds whose energy lie below Ecut are included. Constraints are removed as Ecut is lowered and the rigidity profile of the protein is modified. Rigid cluster decomposition (RCD) plots show the rigidity profile of the protein at each Ecut value which resulted in a change to the flexibility of a Cα atom in the protein backbone. The horizontal axis of an RCD shows the Cα atoms in the primary structure of the protein with flexible regions indicated by thin black lines and rigid regions by thick coloured blocks. The colour associates each residue with a particular cluster and the vertical axis shows the Ecut value used to determine each rigidity profile. Examples of RCD plots can be seen in Figure 2.

As in [2] overall rigidity of the protein is expressed by the function $f_5$, defined in (1). It is determined by how many of the Cα atoms are included in the five largest rigid clusters. The total number of Cα atoms in the protein is $N_{C\alpha}$. Rigid clusters are numbered in order of decreasing size so that $n_i(E_{cut})$ is the number of Cα atoms in rigid cluster $i$ at cutoff value Ecut.

$$f_5(E_{cut}) = \frac{1}{N_{C_\alpha}} \sum_{i=1}^{5} n_i(E_{cut}). \quad (1)$$

Rigidity dilution (RD) graphs plot $f_5$ against Ecut and show how the rigidity of the main chain of the protein is lost as the cutoff value is lowered.

## 3. Results and discussion

Figure 1 shows two representations of the protein corresponding to Ecut values at the beginning 1(a) and towards the end 1(b) of the bond dilution. Rigidity is maintained in two α-helices towards the base of the protein and at the centre where the active site is located. However, at an Ecut of −1.305 kcal/mol, the protein is almost entirely flexible. The same is observed in the structure 1B6J, for which two RCD plots with and without the inhibitor are shown in Figure 2. The regions surrounding residues 50 and 150 of the protein correspond to the β-hairpin flaps which close over the active site of the enzyme. The flaps become flexible much earlier in the bond-breaking process when there is no active-site inhibitor present. This overall trend was observed in all of the structures studied. Figure 3 shows an RD plot of the structure 1B6J with and without an inhibitor attached. In the RD plot, data is recorded in the graph whenever the rigidity of the Cα atoms changes. Only when a Cα atom ceases to be part of one of the five largest rigid clusters in the protein is $f_5$ reduced. This results in the stepwise nature of the plots. With no inhibitor, the $f_5$ value of 1B6J drops faster with decreasing Ecut, although both structures exhibit rapid rigidity loss in agreement with [2].

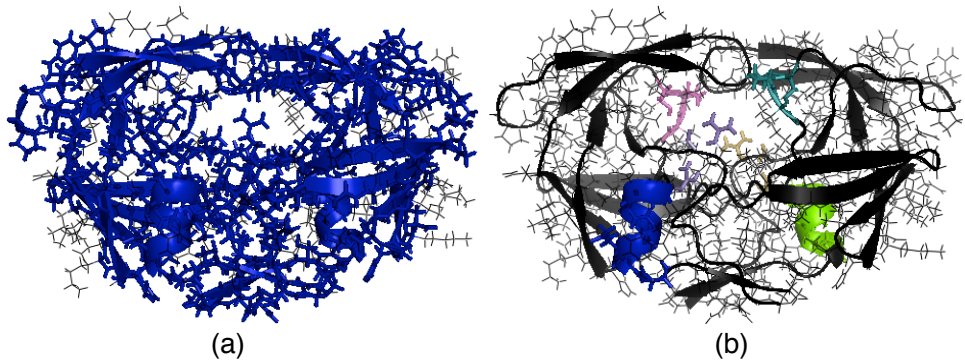

**Figure 1.** PYMOL is used to show the structure 1EC0, its inhibitor having been removed. The protein is largely rigid in (a) with $E_{cut} = 0$ kcal/mol, and flexible in (b), where $E_{cut} = -1.305$ kcal/mol. The secondary structure is shown in a black 'cartoon' representation, and the rigid clusters are shown in different colours for clarity.

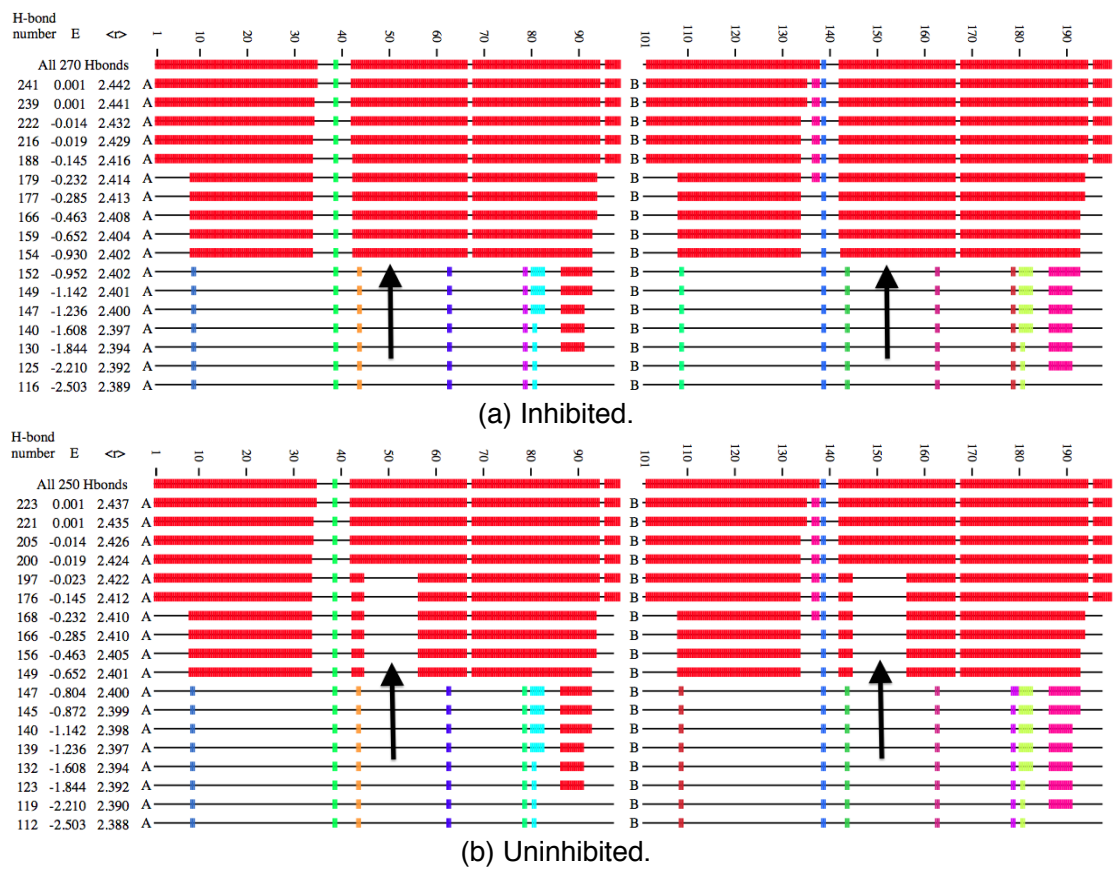

**Figure 2.** The RCD plot for the structure 1B6J is shown in (a), and with its inhibitor having been removed in (b). The thin black lines represent flexible regions of the protein and the thick coloured lines show rigid clusters. The columns on the left hand side of the plot show the number of remaining hydrogen bonds, $E_{cut}$ and the mean coordination value $<r>$, which is an indication of the average number of interactions per atom. The rigidity loss patterns are similar, however the regions indicated by arrows become flexible at higher $E_{cut}$ in (b) than in (a).

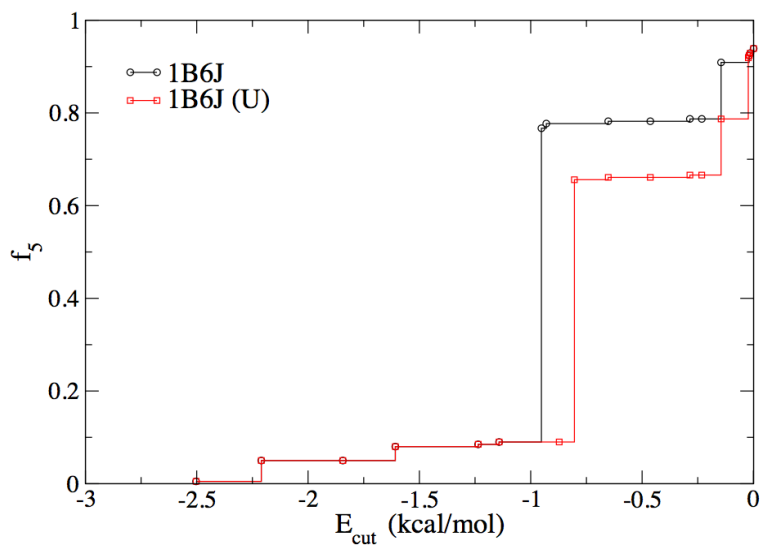

**Figure 3.** The RD plot for the protein with PDB code 1B6J. The suffix (U) signifies the uninhibited copy of the protein. As $E_{cut}$ is lowered, rigidity is lost. Main chain rigidity is lost more rapidly in the uninhibited copy.

## 4. Conclusion and outlook

The overall trend in the rigidity analysis of the 280 crystal structures reveals that the main effect of the inhibitors is to rigidify the flaps of the enzyme which move to permit access to the active site. That the structures studied vary in resolution, inhibitor identity, mutations, and crystallisation conditions suggests that this trend is robust against structural variation. The results tie in with MD simulations such as [9] and experimental evidence [12], where the protein flaps are reported as being mobile in the native state but restricted by the presence of an active-site inhibitor. To expand on the analysis using FIRST, we are currently exploring the use of the mobility software FRODA in combination with elastic network modelling to study predicted protein mobility.